\documentclass[10pt,twocolumn,twoside]{osajnlarc}

\journal{olarch} 
\setboolean{shortarticle}{false}

\ifthenelse{\boolean{shortarticle}}{\colorlet{color2}{color2b}}{\colorlet{color2}{color2}} 

\title{An Upper Bound on the Rate of Information Transfer in Optical Vortex Beams}

\author[]{Matt M. Coles}

\affil[]{School of Chemistry, University of East Anglia, Norwich NR4 7TJ, United Kingdom}

\affil[]{Corresponding e-mail: mattcoles72@gmail.com}

\dates{Compiled \today}

\ociscodes{}

\doi{}

\begin{abstract}

Light endowed with orbital angular momentum, commonly termed optical vortex light, has an azimuthal phase indexed by the orbital quantum number \textit{l}.  In contrast to the two basis states of the optical spin angular momentum, the interest in the information content of optical vortex beams is centred on the assumption that $\lvert{l}\rangle{}$ forms a countably infinite set of basis states. The recent experimental observation that group velocity is inversely proportional to \textit{l} provides a theoretical basis for a practical measure of information transfer. This Letter sets an upper bound on that measure. 
\end{abstract}

\setboolean{displaycopyright}{true}

\begin{document}

\maketitle
\thispagestyle{fancy}

\ifthenelse{\boolean{shortarticle}}{\ifthenelse{\boolean{singlecolumn}}{\abscontentformatted}{\abscontent}}{}

\section{Introduction}

The basic unit of information is the 'bit'; it represents the change in uncertainty from a state of two equally-probable possibilities $(0,1)$ to a definite measurement of the outcome \cite{shannon_mathematical_1948,mackay_information_2003}.  The left- and right- handed photon polarisation basis states, a manifestation of the \emph{spin} angular momentum (SAM), provides a physical realisation of a 'bit'. When states representing bits exist in entangled superpositions, they are known as qubits and are normally depicted as unit vectors in a two-dimensional complex vector space \cite{nielsen_quantum_2000}. Operations on qubits offer the potential for vast improvements over classical computing; for example, the execution of algorithms that run exponentially faster than the best-known equivalent. Despite this, the number of retrievable bits in a quantum system is bounded by that contained in its classical counterpart \cite{holevo_bounds_1973}. A simple corollary is that any theorem that limits classical information content is directly applicable to quantum information. Information can be encoded into a single photon using any measurable degree of freedom \cite{wang_advances_2016}, including: frequency \cite{liu_information_2012,zavatta_manipulating_2014}, spatial structure \cite{tentrup_transmitting_2017, dixon_quantum_2012}, complex polarisation \cite{zhan_cylindrical_2009}, temporal structuring \cite{gauthier_quantum_2013}. 

Light with a phase that varies with azimuthal angle around the wavevector axis is endowed with orbital angular momentum (OAM) and is commonly termed optical vortex (OV) radiation. These beams have Poynting vectors that spiral around the direction of propagation \cite{allen_orbital_1992} and have an azimuthally varying phase factor  $e^{-il \phi},$ where $l$ is the OAM quantum number and may be a positive or negative integer. In some situations, the effect of SAM and OAM can be equivalent \cite{simpson_mechanical_1997}, but in general they have different physical manifestations. In randomly oriented samples, SAM is responsible for differential interactions with chiral molecules \cite{cameron_optical_2012, coles_chirality_2012, loffler_circular_2011, zhao_enantioselective_2016}. However, the enantiomer-dependent electric quadrupole transition moments can engage with OAM to produce some chiral effects \cite{brullot_resolving_2016}. The recent interest in optical vortex light is due to the large number of potential, and realised, applications. For example: detecting the rotary or lateral motion of particles in the beam cross-section \cite{milione_remotely_2017,rosales-guzman_experimental_2013, cvijetic_detecting_2015}, the masking of parent stars with an OV coronograph to allow direct imaging of companion objects \cite{swartzlander_jr._astronomical_2008,serabyn_w._2017}, or imposing one form of optical torque on a nano- or micro- scale particle \cite{bradshaw_manipulating_2017}.
The interest in the information content of OV photons stems from the assumption that, since $\lvert{l}\rangle{}$ forms a countably infinite set of basis states, the number of bits encoded in a photon is only bounded by experimental effects. Specifically, the question of whether there is a maximum information capacity for OAM light is an active area of research \cite{chen_is_2016,potocek_quantum_2015}. The study of \emph{singularimetry} suggests a practical limit on the use of OAM beams for information transfer, since interaction with a topological aberration results in decomposition of an optical vortex into multiple lower-order beams \cite{dennis_topological_2012}. However, the benefits of OAM beams for eavesdropping-resistant free-space information transfer \cite{gibson_free-space_2004} and optical fibre to free-space coupling with artificial turbulence \cite{jurado-navas_850-nm_2015} have been experimentally verified. It has been demonstrated that four light beams with different values of orbital angular momentum can be multiplexed and demultiplexed allowing transmission of over one terabit per second \cite{wang_terabit_2012}. The issue of single photon detection is an active area of study \cite{hadfield_single-photon_2009} as quantum information processing is dependant on the entanglement of pairs of  \cite{mair_entanglement_2001} (or multiple \cite{hiesmayr_observation_2016}) OAM photons.

\section{Modification of the phase and group velocity of light}
The constraint typically applied when modelling a electromagnetic field in a waveguide is that the field must be zero at the boundaries.  The consequences are that the axial wavevector becomes dependent on the frequency, $\omega$, and the width of the guide \cite{feynman_feynman_2011}, and below a certain critical frequency, the field decays exponentially in the guide. Dispersion may also be caused by other geometric boundary conditions or by interaction with a medium, and through either constraint the phase and group velocity become separate. The latter is the velocity at which the envelope of a wave packet travels through space. Here, the geometric boundary conditions are enforced by the lights spatial structure. As structured light has a transverse component of the wavevector, the path length has an additional contribution. To secure a result that is \emph{z}-independent the paraxial approximation is required, and the group velocity for an arbitrary optical field becomes:
\begin{equation}
    \label{group}
   v_g= \frac{c}{1 +\langle {\hat{k}_{\perp}}^2 \rangle / 2k^2_0},
\end{equation}
where $\hat{k}_{\perp}=-i\nabla_{\perp}$ and $k_0$ is the magnitude of the wavevector. Any structured beam with a non-zero expectation value for $\hat{k}_{\perp}$ will experience $v_g<c$. Here we consider a Laguerre-Gaussian (LG) optical field as representative of an OV. However Bessel beams, also endowed with OAM, have similarly been experimentally verified as having a subluminal velocity \cite{hu_ultrashort_2002, alfano_slowing_2016}. Using the LG optical field profile delivers the expectation value as:
\begin{equation} 
    \label{perp}
   \langle {\hat{k}_{\perp}}^2 \rangle = \frac{2}{w^2_0} \big(2p + \lvert l \rvert + 	1 \big),
\end{equation}
where $w_0$ is the minimum beam waist and $p$ is the radial quantum number \cite{bareza_subluminal_2016}. The OAM photon (group) velocity is then inversely proportional to $l$. The integer $p$ is a measurable quantised degree of freedom for a photon, can encode information \cite{karimi_exploring_2014} and plays a role in angular beam shifts \cite{hermosa_radial_2012}. For clarity, the proceeding analysis assumes $p=0$ and $l \ge 0$. The inclusion of the radial and negative $l$ modes will be discussed at the end of this Letter. 

\section{Limits on Photon Information Conveyance}
Theoretically the number of symbols, $N$, required to communicate $B$ bits of information is $2^B$, assuming an equal probability of each symbol occurring \cite{mackay_information_2003}. Let us take an example: Using a basis of $l \in \{0,1\}$ conveys one bit per photon. The addition of states $l$=2 and 3 provides four total outcomes and therefore the possibility of encoding two bits of information. Visualising three photons each with two possible states (for example, left and right circular polarisation) provides justification for the requirement of eight possible outcomes to encode three bits. In the case of OAM photons, the encoding of three bits of information requires the ability to perfectly distinguish between the first eight $l$ values. Extending this argument and including the  Gaussian ($l=0$) mode,  a measured state with $l+1$ possible symbols conveys $N=log_{2} \big(  l  + 1 \big)$ bits of information. The exponential increase in required states is well-known in spatial encoding: to encode 10.5 bits per photon Tendrup et al \cite{tentrup_transmitting_2017} needed to employ 8 times more symbols than previous work, which reported 7 bit per photon as highest value for random keys \cite{dixon_quantum_2012}, and is comparable to what has been achieved in temporal and polarization encoding \cite{gauthier_quantum_2013}.	

To elucidate the issues surrounding detection, we consider a device that will either generate a photon with the phase structure of an $l=1$ or $2$ LG optical vortex ($p=0$). The detection of multiple photons allows for the resolution of either of the donut modes and a precise determination of whether $l=1$ or $2$ light is being sent. However, since the intensity distributions of each mode overlap considerably, the detection of a \emph{single} photon will not provide enough information to reduce the uncertainty in $l$ 	to zero, and will therefore convey only a fraction of a bit. One might use two distant $l$ values, for example $l=1$ and $l=20$, so that the overlap of the intensity distributions is essentially zero; the detection of a single photon would then spatially resolve the OAM modes, within a high confidence interval. It is worth noting that this method requires increasingly large gaps between $l$-values since the overlap between consecutive modes increases as $l \rightarrow \infty $. It has been shown that a Mach-Zehnder interferometer with a rotated Dove prism in each light-path can form the base units of a device that can, in principle, distinguish an arbitrary large number of OAM states at the individual photon level \cite{leach_measuring_2002}. Thus, to proceed we assume that photon OAM states are perfectly differentiable. In fact, high-fidelity OAM sorting may be possible with nano-scale detectors \cite{mei_-chip_2016}.

The assumption here is that individual photons and the contained information travels at the free-space group velocity. In a medium, information can propagate faster than $v_g$ (but not faster than \emph{c}) \cite{stenner_fast_2005}, however it is established that individual photons travel at the group velocity \cite{steinberg_dispersion_1992,giovannini_spatially_2015}. Indeed, it is hard to imagine a situation where optical information arrives before the detection a photon.

\begin{figure}[htbp]
\centering
\fbox{\includegraphics[width=\linewidth]{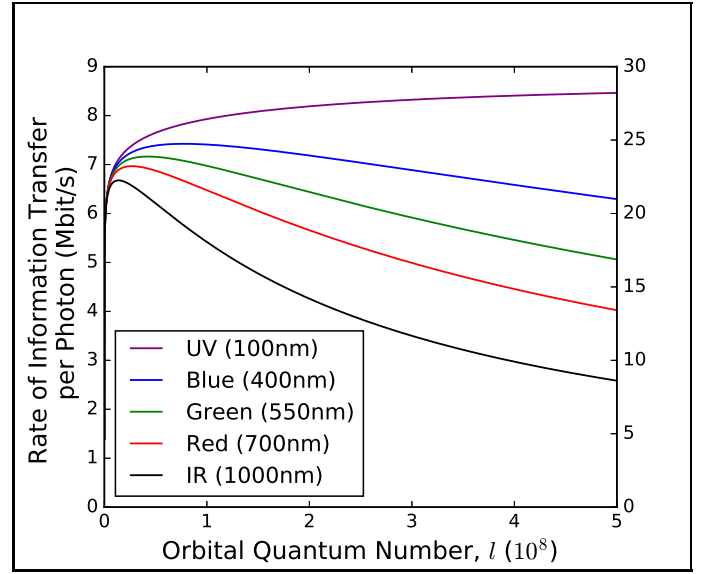}}
\caption{A plot of the rate of information transfer, $\mathcal{I}$, against the distinguishable orbital angular moment values, $l$, for a range of near-visible wavelengths. The beam-waist is fixed at 2.3mm and transfer is considered over a distance of 1km. The right-hand scale is in units of 299,792 bits/s - the rate of information transfer over 1km of a photon travelling at $c$ and carrying one bit. The UV example has a maximum value beyond the range of the graph of $l \approx 10^9$.}
\label{fig1}
\end{figure}

To proceed we consider a photon travelling at the group velocity over a distance $d$ to a detector; the time taken is $\delta t =d/v_g$ and the rate that information arrives per photon per unit time is delivered as:
\begin{equation}
    \label{info}
   \mathcal{I} = \frac{c}{d} \frac{log_{2}\big(  l + 1\big)}{ 1 + \big( \frac{1}{k_{0}w_{0}}\big)^2 \big(  l  + 1\big) },
\end{equation}
where $d$ is the distance travelled. Finding where the derivative with respect to $l$ is zero yields the maximum value of the rate of information transfer:
\begin{equation}
    \label{max}
   \mathcal{I}_{max} = \frac{c}{d} \frac{W\big( \frac{k_{0}^2 w^2_{0}}{e} \big)}{ln\big(2\big) }, \quad  l_{max}=e^{1+W\big( \frac{k_{0}^2 w^2_{0}}{e} \big)}-1, 
\end{equation}
where $c$ is the speed of light, $d$ is the distance travelled.  Here, $W$ is the principal branch of the Lambert-W function \cite{euler_serie_1783}, which is the inverse of $f\big(W \big)=We^{W}$. These are the main results of this Letter. Figure \ref{fig1} displays a plot of $\mathcal{I}$ against $l$ based on Equation \ref{info} over a transmission distance of 1 km. For an infrared photon $\mathcal{I}_{max}$ is obtained when $l \approx 1.38\times10^7$; this corresponds to a transfer rate of 6.68 megabits per photon per second, which is $\approx 22$ times the information transfer rate of a two-state photon travelling at $c$, here designated by the constant $\mathcal{I}_{2S}=c/d$. Dividing Equations \ref{info} and \ref{max} by $\mathcal{I}_{2S}$ delivers results that are manifestly independent of distance. Specifically, Equation \ref{info} becomes the proportionate increase (or decrease) in information conveyance afforded by using OAM basis states over another two-state photon degree of freedom, e.g. polarisation. 

In the limit $l \rightarrow \infty$, similarly $\mathcal{I} \rightarrow 0$ as the light slows to a halt, thus all values of $k_0$ have a maximum value. The UV example in Figure \ref{fig1} has a maximum value beyond the range of the graph of $l \approx 10^9$, where $\mathcal{I}_{max}=28.6 \times \mathcal{I}_{2S}$. Choosing an experimentally realisable OAM content of $l=300$ \cite{fickler_quantum_2012} for a photon with $\lambda =400$nm delivers an information content of $8.2 \times \mathcal{I}_{2S}$. In the latter example, an approximate three-fold increase in the information capacity to  $\mathcal{I}_{max}=23.9 \times \mathcal{I}_{2S}$ requires more than $4.2 \times 10^7$ extra distinguishable $l$-values. 

A curious observation is that the imaginary parts of the analytic continuation of the Lambert-W function describes a surface with striking similarity to a plane of constant phase in an optical vortex. 

\section{Conclusion}
Often discussed in the recent studies on optical OAM states of high dimensionality is the potential for large photonic information capacity \cite{malik_direct_2014,shi_scan-free_2015}. This Letter sets out an analytically tractable upper bound on information conveyance for photons endowed with OAM. The conclusion is clear: it becomes progressively more difficult to encode information in the OAM degree of freedom of a photon as $l$ approaches $l_{max}$, given in Equation \ref{max}. For clarity, the arguments presented above ignore the radial quantum number, $p$ and values of $l<0$. However, Equation \ref{info} is easily modified by the replacement $  l \rightarrow  2\lvert l \rvert+  p $ in the numerator and $  l \rightarrow \lvert l \rvert+  2p $ in the denominator. It is worth noting that the theory presented here applies only to information conveyance; \emph{stationary} light pulses \cite{liu_observation_2001} may still encode a theoretically unbounded amount of information - limited only by the required exponential increase in detectable basis states. Thus, there are less implications for OAM beam use as optical memory \cite{veissier_reversible_2013}. In fact, the difference in group velocity can aid with separation of possible OAM states in the temporal domain. Holevo's theorem \cite{holevo_bounds_1973} states that no more than $n$ bits can be obtained from $n$ qubits. Thus, although the results of this Letter have been framed in terms of classical bits, they are equally applicable to quantum information transfer. In the special case of superdense coding, at most $2n$ bits may be retrieved from $n$ qubits and Equations \ref{info} and \ref{max} are multiplied by a factor of two \cite{bennett_communication_1992, williams_superdense_2017}.

\bigskip
The author gratefully acknowledges helpful insights and comments from Professor David L. Andrews and Dr David S. Bradshaw. 

\bigskip

\bibliography{ollib}

\ifthenelse{\equal{\journalref}{ol}}{%
\clearpage
\bibliographyfullrefs{ollib}
}{}

\end{document}